\documentclass[conference]{IEEEtran}
\usepackage[utf8]{inputenc}
\usepackage[T1]{fontenc}

\usepackage{cite}
\usepackage{amsmath,amssymb,amsfonts}
\usepackage{algorithmic}
\usepackage{textcomp}
\usepackage{xcolor}

\usepackage{graphicx}
\usepackage{paralist}
\usepackage{booktabs}
\usepackage{flushend}
\usepackage{tabularx}
\usepackage{subcaption}
\usepackage{refcount}



\usepackage[english]{babel}

\usepackage[colorlinks=true, allcolors=blue]{hyperref}

\title{Description and Comparative Analysis of \textsc{QuRE}: \\A New Industrial Requirements Quality Dataset}

\usepackage{tikz}
\newcommand\copyrighttext{%
  \footnotesize \textcopyright~2025 IEEE. Personal use of this material is permitted. Permission from IEEE must be obtained for all other uses, in any current or future media, including reprinting/republishing this material for advertising or promotional purposes, creating new collective works, for resale or redistribution to servers or lists, or reuse of any copyrighted component of this work in other works.}
\newcommand\copyrightnotice{%
\begin{tikzpicture}[remember picture,overlay]
\node[anchor=south,yshift=10pt] at (current page.south) {\fbox{\parbox{\dimexpr\textwidth-\fboxsep-\fboxrule\relax}{\copyrighttext}}};
\end{tikzpicture}%
}

\begin{document}

\author{
\IEEEauthorblockN{Henning Femmer}
\IEEEauthorblockA{South Westphalia University \\of Applied Sciences\\ 
Hagen, Germany\\
femmer.henning@fh-swf.de \vspace{-1em}
}
\and\IEEEauthorblockN{Frank Houdek}
\IEEEauthorblockA{Mercedes-Benz AG\\
Sindelfingen, Germany\\
frank.houdek@mercedes-benz.com
}
\and\IEEEauthorblockN{Max Unterbusch, Andreas Vogelsang}
\IEEEauthorblockA{
paluno -- The Ruhr Institute for Software Technology\\
University of Duisburg-Essen \\ Essen, Germany \\ 
\{max.unterbusch, andreas.vogelsang\}@uni-due.de
}
}
\maketitle
\copyrightnotice
\IEEEpeerreviewmaketitle              % typeset the title of the contribution

\begin{abstract}
Requirements quality is central to successful software and systems engineering. Empirical research on quality defects in natural language requirements relies heavily on datasets, ideally as realistic and representative as possible. However, such datasets are often inaccessible, small, or lack sufficient detail. 
This paper introduces \textsc{QuRE} (Quality in Requirements), a new dataset comprising 2,111 industrial requirements that have been annotated through a real-world review process. Previously used for over five years as part of an industrial contract, this dataset is now being released to the research community. In this work, we furthermore provide descriptive statistics on the dataset, including measures such as lexical diversity and readability, and compare it to existing requirements datasets and synthetically generated requirements. 
In contrast to synthetic datasets, \textsc{QuRE} is linguistically similar to existing ones. However, this dataset comes with a detailed context description, and its labels have been created and used systematically and extensively in an industrial context over a period of close to a decade.
Our goal is to foster transparency, comparability, and empirical rigor by supporting the development of a common gold standard for requirements quality datasets. This, in turn, will enable more sound and collaborative research efforts in the field.
\end{abstract}


\section{Introduction}
The quality of natural language (NL) requirements is important for efficient and effective software and systems engineering~\cite{Femmer2019,Mund2015}. 
Deficiencies such as ambiguity, inconsistency, and incompleteness in requirements can lead to significant challenges, including increased development costs and project delays~\cite{Fernndez2016}. 
Recognizing these challenges, the research community has extensively explored methods to detect and mitigate quality issues in NL requirements~\cite{Montgomery2022}. For instance, the concept of ``requirements smells'' has been introduced to identify patterns indicative of potential quality problems, facilitating rapid quality assurance processes~\cite{Femmer17}.

Despite the substantial body of research, publicly available datasets that provide NL requirements annotated with quality-related labels are scarce. Many existing datasets are inaccessible, limited in size, lack rigorous annotation standards, or do not reflect real-world industrial settings. This limitation prevents researchers from validating findings, comparing methodologies, and developing data-driven tools for quality assessment. Moreover, the absence of standardized datasets impedes the advancement of empirical studies aimed at understanding and improving the quality of requirements~\cite{Montgomery2022}.

To address this gap, we present \textsc{QuRE} (\textbf{Qu}ality in \textbf{Re}quirements), a comprehensive dataset consisting of 2,111 NL requirements sampled from real-world automotive projects. 
Each requirement is annotated with quality defect labels derived from an industrial review process, ensuring authenticity and practical relevance. 
This dataset is the largest publicly available resource of its kind and offers valuable insights for both academic research and industrial applications. In addition to releasing the dataset, we conduct a comparative analysis with existing datasets, examining linguistic features such as lexical diversity and readability. We also provide an in-depth discussion on the labeling taxonomy, dataset provenance, and annotation quality, contributing to a more standardized and empirical approach to requirements quality assessment.

By making this dataset publicly available, we aim to support empirical research on the quality of natural language requirements. It provides a solid foundation for benchmarking automated quality assessment tools, training machine learning models, and conducting replication studies. In addition, the real-world industrial context of the data enhances its validity, offering researchers and practitioners a more realistic basis to evaluate techniques and understand quality defects. Ultimately, this contribution supports the development of more reliable, scalable, and context-sensitive approaches to improving the quality of requirements across diverse domains.

The \textsc{QuRE} dataset and all corresponding analysis code are publicly available through our Zenodo project\footnote{\url{https://doi.org/10.5281/zenodo.15656471}}. 

\section{Related Work}
\label{sec:related-work}

%\textbf{Datasets in Software Engineering:}  
Datasets are important in empirical research within SE, particularly in the development and evaluation of automated tools and methodologies. A recent study analyzing 2,196 papers from leading SE venues (ICSE, FSE, ASE, ISSTA) between 2017 and 2022 found that approximately 68\% included research artifacts, with datasets forming a substantial component~\cite{Liu2024}. These findings underscore the field’s growing emphasis on transparency, reproducibility, and data-driven experimentation.

\textbf{Datasets in Requirements Engineering:} Although empirical research in RE is well-established, large publicly available datasets remain relatively scarce. A key challenge is that requirements artifacts are often absent from open-source software repositories, and those from industrial software are rarely released due to confidentiality or proprietary concerns. However, several efforts have been made to collect, curate, and publish requirements datasets. The RE Open Data initiative\footnote{\label{re-open-data}\url{https://zenodo.org/communities/re-data/records}} aims to centralize these resources.
Currently, the initiative has collected five datasets:
\begin{itemize}
    \item \textbf{PURE (PUblic REquirements Dataset)}: A collection of 79 publicly available requirements documents comprising 34,268 sentences~\cite{ferrari2017pure}. The dataset comes without any description or knowledge about the context of the projects. 
    \item \textbf{PROMISE}: Initially containing 625 labeled requirements (255 functional, 370 non-functional), PROMISE has been widely used in requirements classification research. PROMISE\_exp, an expanded version, includes 444 functional and 525 non-functional requirements, curated through expert consensus~\cite{lima2019}. Further enhancements by Dalpiaz~et~al.~\cite{Dalpiaz2019} introduced multi-label annotations for both functional and non-functional requirements.
    \item \textbf{Intralogistics Requirements}: A 13-page requirements document for a warehouse management system compiled from multiple industrial projects, anonymized to remove sensitive information~\cite{Intralog-data}.
    \item \textbf{User Stories}: A collection of 22 sets of 50+ user stories each from various open projects, accompanied by references to their original repositories and issue trackers~\cite{US-data}.
    \item \textbf{Security Requirements}: Three datasets consisting of requirements for three different systems, each labeled to indicate whether it addresses security concerns~\cite{Security-data}.
\end{itemize}

\textbf{Datasets in Requirements Quality Research:}  
Frattini~et~al.~\cite{frattini2024requirements} introduced an extensible ontology that captures research on requirements quality, which includes a curated list of datasets used in this domain.\footnote{\url{http://reqfactoront.com/content/datasets}} According to the ontology, 57 datasets have been used across 34 publications. These datasets vary widely in size, ranging from 1 to 26,829 requirements, with an average of 943 and a median of 59 requirements.
The authors report that only 10 out of the 57 datasets are publicly accessible---either as open access or via a direct link in the paper. An additional 3 are marked as available ``upon request''. Among these 13 non-proprietary datasets, the number of requirements ranges from 4 to 600, with an average of 137 and a median of 20. Notably, 7 of the 13 were sourced from industry practitioners.

These observations point to significant limitations in the availability and quality of datasets for requirements quality research:
\begin{itemize}
    \item A majority of datasets (44 out of 57) are inaccessible to the research community.
    \item The datasets that are publicly accessible tend to be small in scale (median of 20 requirements).
    \item Although 42 out of 57 datasets originate from practitioners, only 7 are annotated or labeled by practitioners.
\end{itemize}

Due to these constraints, some researchers have resorted to generating synthetic requirements data. For example, Fantechi~et~al.~\cite{fantechiInconsistencyDetectionNatural2023} used ChatGPT to generate conflicting requirements based on a seed set and subsequently asked the model to detect the inconsistencies. The degree of representativeness of such synthetic data and, consequently, the validity of studies executed with these datasets remains to be studied.

\section{The \textsc{QuRE} Dataset}
The main contribution of this paper is to provide a dataset for requirements quality analyses. In our view, it is central to understand the context of this dataset to be able to analyze its quality. Therefore, this section explains the history of this dataset through its individual phases and describes its structure.

\subsection{Context and History} \label{sec:context}

The dataset consists of requirements that originate from specifications of Mercedes-Benz, a large automotive OEM.
Before we published the data set, the data went through several steps that are important to understand for valid research designs. 

\textbf{Step 1 (Context of the requirements and how they have been authored):}
A modern Mercedes-Benz car contains several hundred electronic devices with software running on them (e.g., electronic control units, sensors, and actuators). The specification for a single device typically ranges between 100 and 2,000 pages of requirements, represented mostly as natural language text. Additionally, a device specification refers to 30--300 related documents. The requirements that form the basis for the \textsc{QuRE} dataset are extracted from these specifications. They were written by authors with different roles, such as component engineers or technological experts. Boutkova~\cite{boutkova2011experience} and Krisch~et~al.~\cite{krisch2017sprachliche} provide a good overview of the context of these requirements.

\textbf{Step 2 (Original sampling of dataset):}
The data was first leveraged, but not published, as part of Krisch's PhD thesis~\cite{krisch2017sprachliche}.
The motivation of Krisch's underlying research~\cite{krisch2017sprachliche,Landhausser15,krisch-etal-2016-lexical} was to create rules to identify quality defects in natural language requirements, in particular from a tester’s perspective. The categories of problem candidates (now often referred to as \textit{requirements smells}~\cite{Femmer17}) included weak words, passive voice, or ambiguous references~\cite{krisch2015myth}.  
 
To build a reliable weak word classification, Krisch used a Mercedes-Benz catalog of 103 weak words, analyzed more than 100 specification documents, and identified more than 100,000 requirements containing these weak words. In her work, she identified linguistic rules that separated the problematic usage of a weak word from the unproblematic~\cite{krisch2017sprachliche}.
 
From the reduced dataset, a subset was created for regression testing of automated requirements smell detection tools.
The criterion for selection was that every weak word from the Mercedes-Benz catalog with sufficient usage in real-world specifications shall appear, both in problematic and unproblematic requirements. Labeling was performed by up to three industry experts, as described in Section~\ref{sec:dataset_description}.
 
              % Please note the following properties of the sample requirements:
The requirements analyzed were intentionally sampled from different maturity stages. Some of the requirements might originate from 'in-draft' specifications, others from finalized specifications.
The selection of the requirements was driven by the purpose of having a dataset for regression testing of smell detection approaches. 
%To publish the requirements, we further anonymized the requirements and blinded for instance model names, company names and some more details. From a requirements evaluation perspective, however, the requirements are untouched.

From this, Krisch created an English and a German requirements set, from which this paper only releases the English version due to the broader applicability in the community.

\textbf{Step 3 (Usage of the data in an SLA to Qualicen):}
In 2018, shortly after Krisch finished her PhD, Daimler searched for a new contractor for maintaining and extending the existing analyses. Qualicen, a startup founded in 2015, had previously released a tool, the Qualicen Scout~\cite{femmer2017,femmer2018requirements}, which had shown the most promising results of all competitors according to Krisch's analysis. As part of long-term contracts between the companies, Qualicen signed a service level agreement~(SLA) that promised that the analyses would not deteriorate by a certain level in precision and recall over the original results. The dataset was used in this regard for the following five years.

\textbf{Step 4 (Final preprocessing by us before release):}
After this contract has ended, we want to enable the community to also profit from this real-world dataset, which has been extensively used in the previous years. To do so, we had to anonymize the requirements to some extent by removing car model names, company names, and some technical details that are irrelevant from a requirements evaluation perspective.
In addition, we cleaned the dataset by excluding ambiguous data. This included 31 rows with inconsistent data (a human identified a defect although there was no weak work in the requirement) and 33 rows where the human labeler indicated that they could not decide whether the requirement had a defect or not. In total, we excluded 64 rows ($<3\%$).

\subsection{Dataset Description}\label{sec:dataset_description}

The released \textsc{QuRE} dataset contains 2,111 unique requirements in natural, unconstrained English. Table~\ref{tab:dataset_extract} shows an extract of the dataset. We provide the dataset as a CSV file with four columns: 
\begin{itemize}
    \item \textbf{id}: A unique number identifying the data point.
    \item \textbf{requirement}: The requirement text as it is represented in the original specification.
    \item \textbf{weak\_word}: A label representing a weak word apparent in the requirement. The selection of weak words originates from the dissertation of Krisch~\cite{krisch2017sprachliche}. Weak words signal potential defects in requirements. Labels are umbrella categories to account for different linguistic deviations (i.e.\ inflections) of a weak word. For example, the label \texttt{bad} subsumes the words ``worse'' and ``worst''.   
    \item \textbf{defect}: A human-labeled assessment of whether the corresponding weak word in the requirement constitutes a quality defect (value \texttt{defect}) or not (\texttt{ok}). For example, in the requirement ``\textit{If the button is pressed for a long time, then ...}'', the word ``long'' is problematic, whereas in ``\textit{The long range radar shall send ...}'', the same word is appropriate in that context and, therefore, not considered a defect.
    The defect column was labeled by Mercedes-Benz employees, specifically software testers. If individual cases were unclear, the labels were negotiated by up to three experts. Unfortunately, we could not retrieve any information on the consistency of the experts' labeling, such as inter-rater agreement metrics.
\end{itemize}

\begin{table*}
    \centering
    \caption{Extract from the dataset}
    \label{tab:dataset_extract}
    \begin{tabularx}{\textwidth}{@{}rXll@{}}
    \toprule
    \textbf{id} & \textbf{requirement} & \textbf{weak\_word} & \textbf{defect} \\
    \midrule
       10  & The Contractor shall ensure that adequate test resources are available for every market-specific version of the component. & adequate & ok \\
     16 & In case of abnormal condition, the system shall provide adequate information to driver and/or service personnel. & adequate & defect \\
       326  & If the value is less than or equal to tbd, the window is closed. & close & ok \\
       454  & The component has to separate stationary and moving objects with same speed on own and neighbour lane close to a row of cars that are stopped or are driving at the same speed. & close & defect \\
       605 & The contractor shall be prepared that the dimensions will change in detail. & detail &defect\\
       668 & For details see ``Technical Specifications for Diagnosis''. & detail & ok\\
       1985 & The contractor shall depict special equipment scopes in the pin assignment. &special &ok\\
       2127 & The component shall provide a special sound setting for speech in the telephone/SDS mode. &special&defect\\
    \bottomrule
    \end{tabularx}
\end{table*}

Most requirements appear only once in the dataset, but some have multiple weak words. In such cases, each weak word is represented and assessed as an independent row.

As the dataset was used both internally at Mercedes-Benz as a benchmark dataset and externally as a basis for contractual agreements with partner companies, the annotations have industry quality.

\begin{figure}
    \centering
    \includegraphics[width=\linewidth]{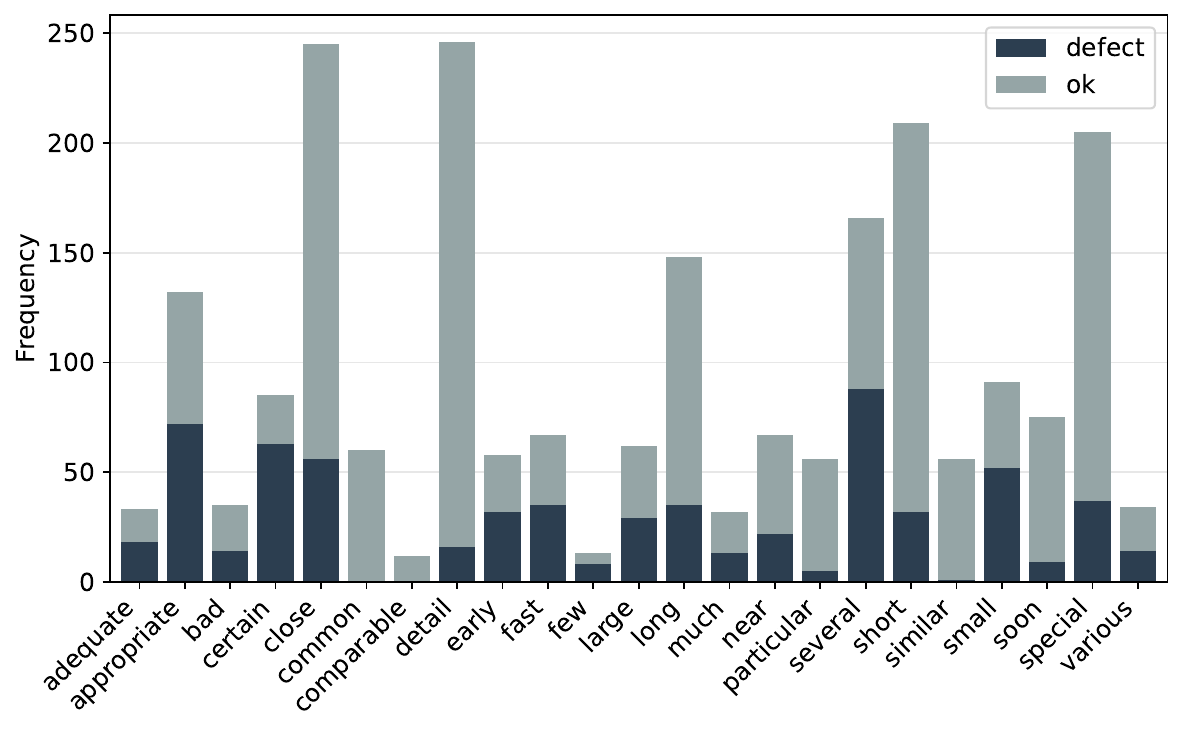}
    \caption{Distribution of weak words in \textsc{QuRE}}
    \label{fig:defect_distribution}
\end{figure}

The weak word column contains 23 unique labels. The most frequent labels are ``detail'' (11.2\%), ``close'' (11.2\%), and ``short'' (9.6\%).
A total of 633 requirements, almost a third, are labeled as \texttt{defect} while 1,478 requirements are labeled as \texttt{ok} despite the weak word.
Fig.~\ref{fig:defect_distribution} visualizes the distribution of weak word indicators.
Each bar shows the frequency of a weak word label, with stacked segments indicating \texttt{defect} and \texttt{ok} assessment.

\section{Comparing Existing Datasets in RE Quality Research}

To analyze the generality of our datasets, we compare the linguistic characteristics of our dataset to other generic and requirements-specific datasets.

\subsection{Data Selection}\label{sec:data_selection}

\textbf{Requirements-specific Datasets}: 
To compare our domain- and company-specific dataset with real-world requirements from diverse domains, we selected four of the five datasets in the RE Open Data Initiative's Zenodo community\footnotemark[\getrefnumber{re-open-data}].
We included the PURE, PROMISE, User Stories, and Security Requirements datasets, but excluded the Intralogistics Requirements dataset due to its small size.

If the data was available in CSV or Excel format, we used it directly.
For PURE, we used a preprocessed subset from Kaggle\footnote{\url{https://www.kaggle.com/datasets/computerscience3/public-requirementspure-dataset}}, as the original PURE dataset is a collection of raw documents in various formats.
This subset contains 10,604 unique sentences, representing approximately one-third of the complete PURE dataset.
To avoid confusion with the full PURE dataset, we refer to our sample as PURE* throughout this work.

\textbf{Synthetic Requirements}:
As described in Section~\ref{sec:related-work}, authors are experimenting with generated synthetic requirements to compensate for the lack of large-scale requirements datasets.
To see how the linguistic character of LLM-generated requirements compares to real-world data, we generated three datasets (named \textit{Basic}, \textit{Context}, and \textit{Context \& Role}) of 500 requirements each.
We used \mbox{gpt-4.1-2025-04-14}, OpenAI's flagship GPT model at the time of writing, with the standard temperature of 1.

For the \textit{Basic} dataset, we executed the user prompt ``\textit{Generate 100 software requirements}'' five times, mimicking a naive approach to synthetic data generation.
The \textit{Context} dataset was created by running the user prompt ``\textit{Generate 100 software requirements for a <system\_name>}'' five times for five different domains (e-commerce platform, hospital management system, autonomous vehicle control system, smart home automation, and financial trading platform).
Finally, the \textit{Context \& Role} dataset used the same user prompt as \textit{Context}, but additionally included a preceding system prompt following the ``role'' (aka. ``persona'') pattern~\cite{white2023prompt}: \textit{You are an experienced requirements engineer in the <domain> domain}'' for each of the five domains.
This progression allows us to assess whether domain-specific context and role-playing patterns produce more realistic-looking synthetic requirements compared to the generic prompt.

The output of the LLM was formatted as Markdown, with headings that sometimes separated requirement type (e.g., functional and non-functional) or grouped related requirements (e.g., user management), along with auxiliary text and ordered lists of the requirements.
To produce a clean CSV file containing only the requirements, the headings, auxiliary text, and numbering were removed.
We also removed identically duplicate requirements, yet did not attempt to de-duplicate requirements that were only semantically similar.

\textbf{Generic English}: Inspired by the PURE paper~\cite{ferrari2017pure}, we also compare the datasets to the Brown corpus~\cite{francis1979brown} to relate the our findings to generic English. The dataset was also retrieved from Kaggle\footnote{\url{https://www.kaggle.com/datasets/nltkdata/brown-corpus/data}} and contains 57,340 sentences originating from 500 samples of diverse genres of English prose.

\subsection{Data Analysis}

We automatically compute and analyze metrics from three dimensions of textual complexity to compare the linguistic properties of the datasets. The metrics were selected based on their ability to provide a first overview of the dataset, building on previous, similar works~\cite{ferrari2017pure}. 

\textbf{Lexical complexity} captures vocabulary richness and information density.
We report the number of tokens (all words excluding punctuation) and lexical words (tokens without stop words or numbers), vocabulary size at different levels of granularity, lexical diversity, and lexical density.
Lexical diversity is calculated as the ratio of stem vocabulary size to total lexical words, where stems represent the morphological roots of lexical words~\cite{ferrari2017pure}.
Lexical density is defined as the ratio of total lexical words to total tokens \cite{brunato2018sentence}.

Measures of \textbf{syntactical complexity} account for the structural sophistication of a text.
We use the average sentence length (in lexical words and tokens), the average number of clauses per sentence~\cite{lu2010automatic}, and average parse tree depth, the longest path from root to leaf in the dependency tree of a sentence~\cite{brunato2018sentence}.

Finally, we used Flesch-Kincaid Grade Level~\cite{kincaid1975derivation} as a \textbf{readability} metric, which refers its scores to the skills of U.S. grade levels. 
The metric is based on the average sentence length in tokens and the average number of syllables per word.
Even though requirements are usually written and read by graduates, we still consider this metric helpful for putting the reading complexity of the sentences into perspective.

\subsection{Results}

\textsc{QuRE} represents the second-largest real-world requirements dataset in this comparison, based on the number of rows, unique sentences, and tokens (Tab.~\ref{tab:dataset_comparison}).
Except PURE*, real-world requirements exhibit consistent lexical diversity.
In contrast, both PURE* and the Brown corpus show noticeably lower diversity.
Synthetic datasets, on the other hand, demonstrate higher lexical diversity and significantly greater lexical density than their real-world counterparts.

\begin{table*}
\caption{Linguistic characteristics of requirements datasets}
\label{tab:dataset_comparison}
\centering
\begin{tabular}{@{}lrrrrrrrrr@{}}
\toprule
 & \multicolumn{5}{c}{\textbf{Real-World Requirements}} & \multicolumn{3}{c}{\textbf{Synthetic Requirements}} &  \\
\cmidrule(lr){2-6} \cmidrule(lr){7-9}
 & \textsc{QuRE} & PURE* & PROMISE & Security & User Stories & Basic & Context & Context-Role & Brown \\
\midrule
\textbf{General}\\
\hspace{1em}Number of Rows & 2,187 & 11,440 & 625 & 511 & 1,680 & 500 & 500 & 500 & 57,340 \\
\hspace{1em}Number of Unique Sentences & 2,111 & 10,604 & 624 & 445 & 1,673 & 494 & 500 & 500 & 56,418 \\
\textbf{Lexical Complexity}\\
\hspace{1em}Number of Tokens & 45,703 & 205,922 & 12,413 & 12,746 & 40,392 & 4,721 & 4,342 & 5,757 & 1,032,154 \\
\hspace{1em}Number of Lexical Words & 26,788 & 114,078 & 7,380 & 7,545 & 19,132 & 3,376 & 3,228 & 4,205 & 530,744 \\
\hspace{1em}Vocabulary Size (Tokens) & 4,852 & 7,994 & 1,671 & 1,354 & 3,328 & 886 & 1,252 & 1,332 & 42,428 \\
\hspace{1em}Vocabulary Size (Lexical Words) & 4,522 & 7,330 & 1,528 & 1,249 & 3,174 & 813 & 1,173 & 1,249 & 41,371 \\
\hspace{1em}Vocabulary Size (Stems) & 3,170 & 4,693 & 1,089 & 845 & 2,065 & 614 & 904 & 945 & 26,061 \\
\hspace{1em}Lexical Diversity & 0.12 & 0.04 & 0.15 & 0.11 & 0.11 & 0.18 & 0.28 & 0.22 & 0.05 \\
\hspace{1em}Lexical Density & 0.59 & 0.55 & 0.59 & 0.59 & 0.47 & 0.72 & 0.74 & 0.73 & 0.51 \\
\textbf{Syntactic Complexity}\\
\hspace{1em}Avg. Sentence Length (Tokens) & 21.65 & 19.42 & 19.89 & 28.64 & 24.14 & 9.56 & 8.68 & 11.51 & 18.29 \\
\hspace{1em}Avg. Sentence Length (Lexical Words) & 12.69 & 10.76 & 11.83 & 16.96 & 11.44 & 6.83 & 6.46 & 8.41 & 9.41 \\
\hspace{1em}Avg. Number of Clauses & 2.14 & 2.17 & 2.34 & 3.08 & 3.91 & 1.36 & 1.22 & 1.41 & 2.36 \\
\hspace{1em}Avg. Parse Tree Depth & 5.89 & 5.72 & 5.45 & 6.49 & 6.16 & 3.49 & 3.60 & 4.17 & 5.08 \\
\textbf{Readability}\\
\hspace{1em}Flesch Kincaid Grade Level & 11.01 & 11.36 & 9.82 & 14.77 & 10.50 & 6.55 & 7.70 & 8.04 & 7.65 \\
\bottomrule
\multicolumn{10}{l}{* Preprocessed Kaggle subset of PURE, as discussed in Sec.~\ref{sec:data_selection}}
\end{tabular}
\end{table*}

Real-world requirements also show substantial syntactic complexity. Sentences tend to be longer, contain more clauses, and have deeper parse tree structures. The Security dataset stands out as the most complex across most syntactic metrics, while User Stories feature the highest clause count. \textsc{QuRE} falls into the middle-ground in terms of syntactical complexity in the real-world datasets.

\begin{figure*}
    \centering
    \begin{subfigure}[b]{0.32\textwidth}
        \centering
        \includegraphics[width=\textwidth]{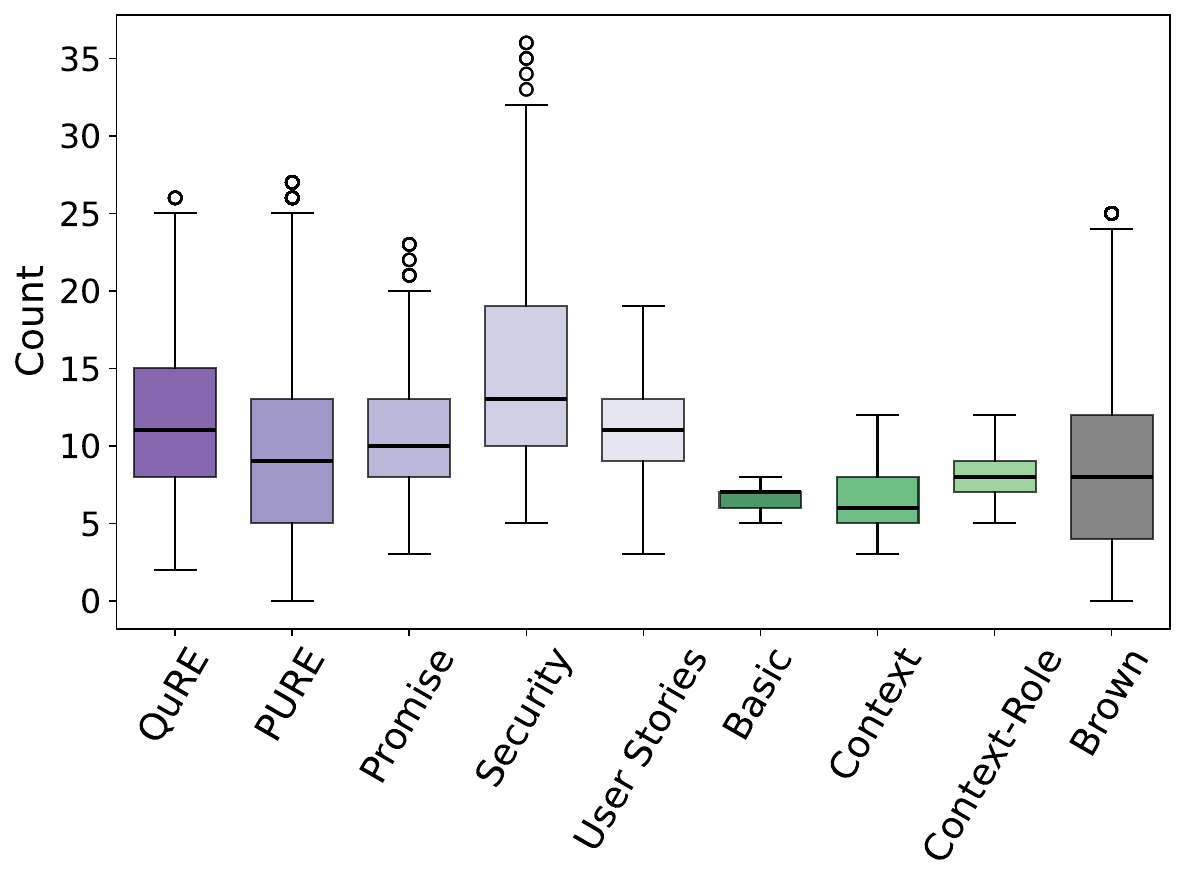}
        \caption{Lexical Words}
        \label{fig:lexical_words}
    \end{subfigure}
    \hfill
    \begin{subfigure}[b]{0.32\textwidth}
        \centering
        \includegraphics[width=\textwidth]{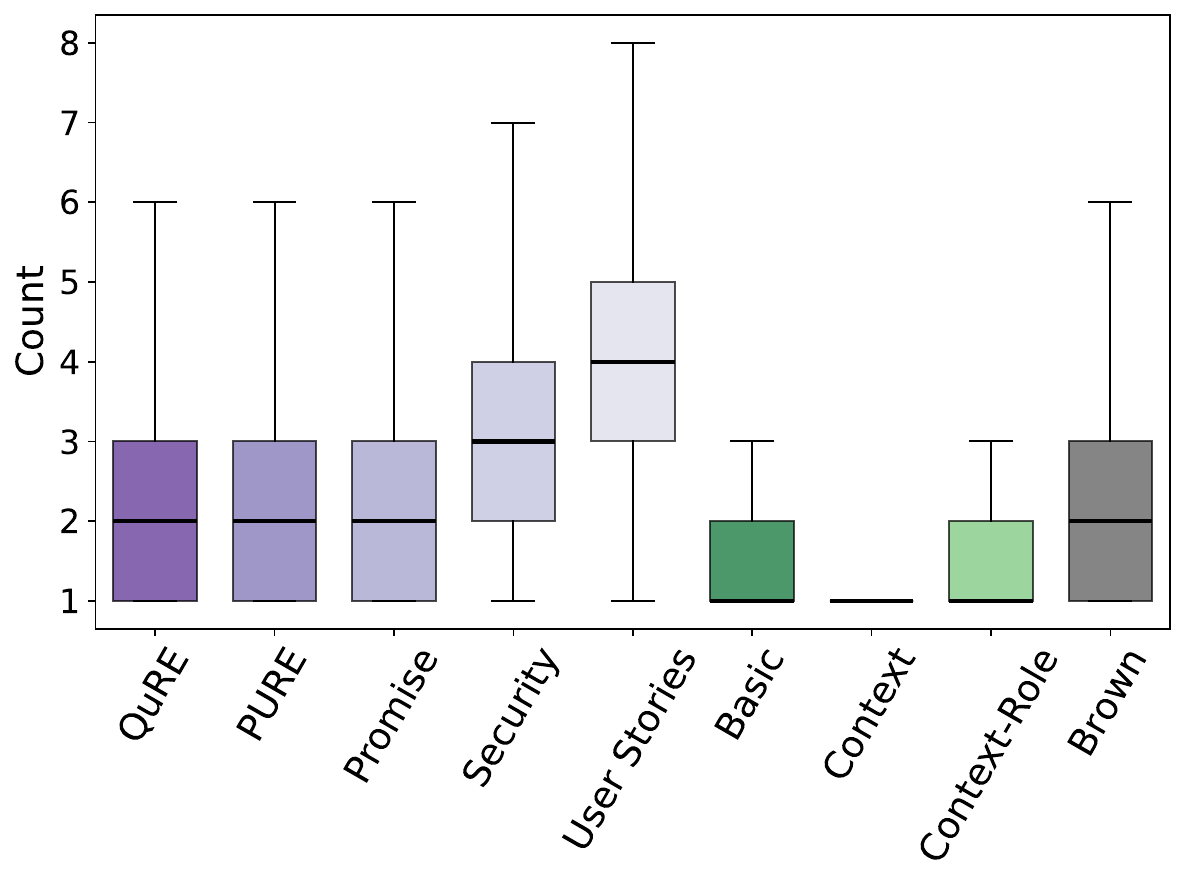}
        \caption{Number of Clauses}
        \label{fig:clause_count}
    \end{subfigure}
    \hfill
    \begin{subfigure}[b]{0.32\textwidth}
        \centering
        \includegraphics[width=\textwidth]{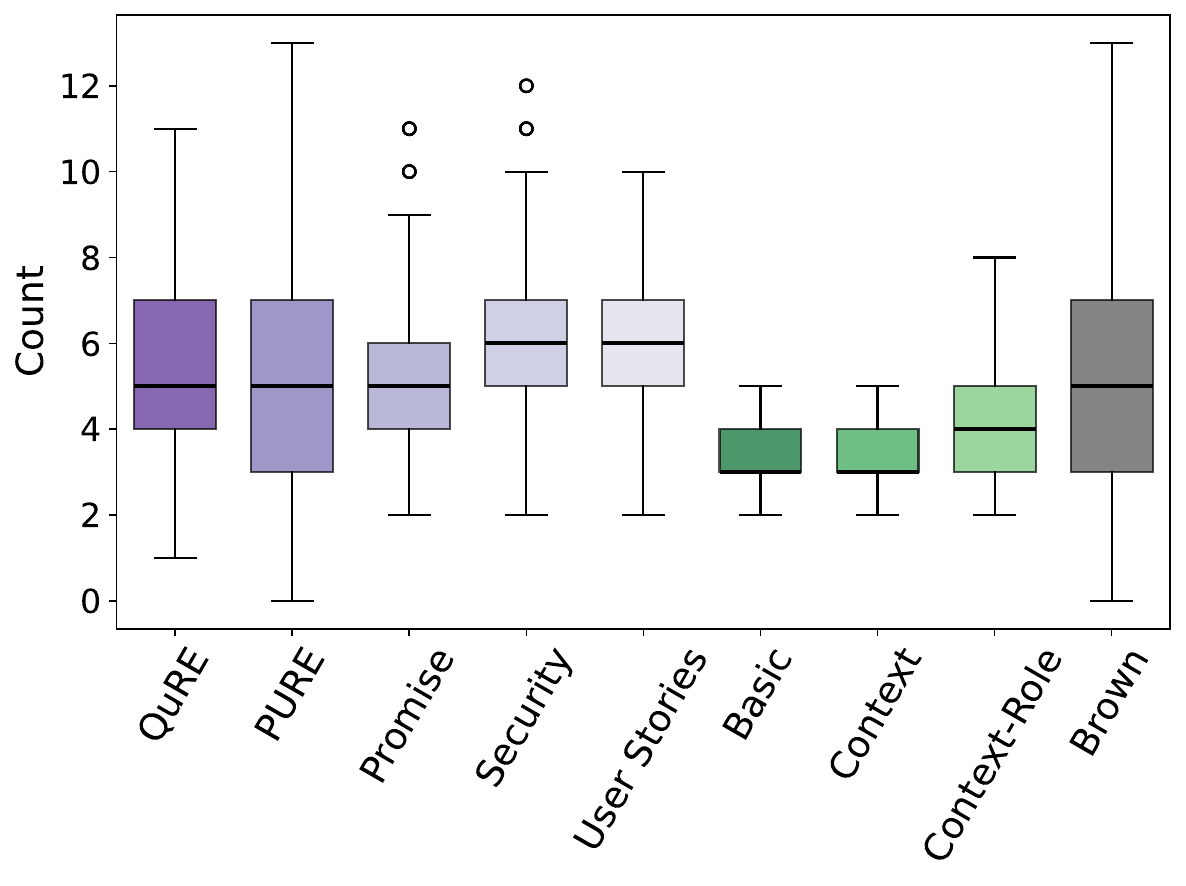}
        \caption{Parse Tree Depth}
        \label{fig:parse_tree_depth}
    \end{subfigure}
    \caption{Distribution of syntactic complexity metrics across sentences}
    \label{fig:syntactic_complexity}
\end{figure*}

In contrast, synthetic requirements are syntactically simpler.
Sentences are shorter, contain fewer clauses, and exhibit shallower parse trees. 
A clear progression emerges with enhanced prompting strategies, from Basic (no context) through Context (project domain) to Context \& Role (domain plus role pattern).
As shown in Fig.~\ref{fig:syntactic_complexity}, this evolution leads to modest increases in complexity, though synthetic texts remain well below the levels observed in real-world datasets.
These datasets also show lower variance, suggesting more uniform sentence structures.

\begin{figure}
    \centering
    \includegraphics[width=\linewidth]{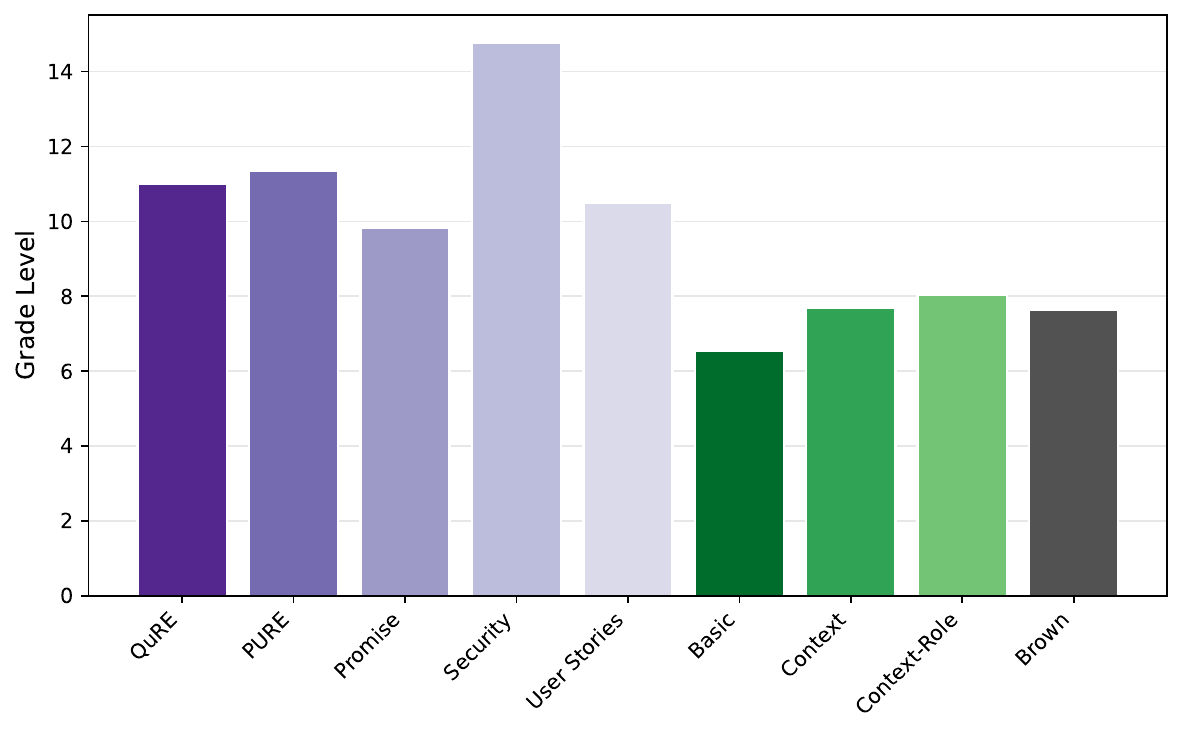}
    \caption{Flesch-Kincaid Grade Level between datasets}
    \label{fig:readability}
\end{figure}

A similar pattern holds for readability.
Based on the Flesch-Kincaid Grade Level metric, real-world requirements tend to be less readable.
Synthetic requirements are generally easier to read and closely match the reading level of the Brown corpus.
Again, \textsc{QuRE} occupies a middle-ground compared to the other real-world datasets (Fig.~\ref{fig:readability}).

\section{Discussion}

We argue that the dataset provides an extensive and realistic view into typical (especially linguistic) defects in RE in practice, as the dataset has served as an internal contract-relevant benchmark in industry for five years. However, due to its context and history (see Sec.~\ref{sec:context}) it comes with reasonable applications and limits thereof, which we want to discuss in this chapter.

\subsection{General Observations}
The \textsc{QuRE} dataset is one of the largest publicly available resources in our comparison. When compared with other real-world datasets, \textsc{QuRE} appears quite typical, not in a negative sense, but rather in that it aligns well with the linguistic characteristics commonly found in other real-world requirements datasets.
On the contrary, what distinguishes \textsc{QuRE} is its clearly described context and documented use in industry.

In our early examination, synthetic requirements often exhibit a noticeably lower degree of realism. Structurally, they tend to be less complex and less nuanced than real-world requirements. One plausible explanation for this is that LLMs are more frequently trained on simplified, textbook-style examples than on full-fledged, domain-specific requirements documents. We do not claim that generating realistic requirements with LLMs is impossible. Rather, our findings suggest that generating representative and realistic synthetic requirements is not straightforward. A naive generation approach risks producing content that lacks the depth and structure of industrial requirements. Addressing the question of whether, and under what conditions, LLMs can be used to generate high-quality, realistic requirements remains a compelling direction for future research.

\subsection{Opportunities for Using the Dataset in Future Research}
The dataset could continue to serve as the minimal standard for future analyses and tools in this direction. The benchmark contains both requirements in later authoring stages, but also requirements in early stages with more defects. We consider this an advantage, since future RE quality tools are hopefully applied during the writing process and not only after all reviews have passed.

The benchmark is labeled good enough for its industry purpose (and again in this regard very realistic), but we quickly realized in our cooperation that our understanding of requirements quality is very context dependent and to some degree subjective~\cite{krisch2015myth, Femmer2019}. Future research could use the existing labels to extend our understanding of linguistic defects in practice and develop and apply better taxonomies of requirements defects (e.g., based on the work of Frattini~et~al.~\cite{frattini2023requirements}).

\subsection{Limitation for Using the Dataset in Future Research}
The dataset is not suited for any sort of distribution and frequency analysis. Due to the sampling strategy, the dataset has a much higher defect density than regular requirements, both at Mercedes-Benz and (to our experience) also elsewhere. The requirements and defects might not reflect the typical distribution of automotive requirements in their final stage. The labels of the data and their quality reflect industrial standards. That means, single labels may be debatable or even incorrect from an academic point of view. However, from an industry perspective, they are well-suited to serve their purpose. 

Since the requirements were created by Mercedes-Benz authors in their respective projects, their requirements writing style reflects the traditional automotive domain in Germany. In our experience, this style can be found across various systems engineering projects in automotive, avionics, defense, healthcare, or related domains. However, studies executed with this dataset might not generalize to, e.g., pure software requirements written in a user-story template. This first analysis reported in this work does not indicate any striking differences, but we recommend taking this to a deeper level in future work.

Furthermore, the dataset focuses on weak words, which represent only one type of quality defect. Weak words belong to the group of lexical smells~\cite{frattini2023requirements}. Further smell categories include syntactical smells (e.g., referential ambiguity~\cite{Ezzini2022}) and semantic smells (e.g., logical inconsistencies). These categories are equally important, but are not reflected in our dataset.

\subsection{Call-for-Action: Community-Driven Repository for Requirements Datasets}

To foster reproducibility, collaboration, and progress in requirements quality research, we propose the establishment of a centralized dataset repository for annotated software requirements. For example, the RE Open Data Initiative\footnotemark[\getrefnumber{re-open-data}] should be extended to serve as a comprehensive, community-driven platform for sharing and evolving requirements datasets.
All previously described requirements datasets, including those used in our study, should be integrated into this central repository. To ensure interoperability and consistency, all requirements should be stored in a common machine-readable format such as CSV, JSON, or ReqIF.

We envision collaborative contributions taking place during community events such as workshops or dedicated sprints, where researchers can jointly contribute to and extend the labeling of the datasets. Platforms such as GitHub (for development and version control) and potentially Kaggle (for public challenges or benchmarking tasks) may serve as complementary dissemination channels.
In addition to the raw data, all analysis workflows should be published as Jupyter notebooks within the same repository. This will allow for transparent and replicable experiments, facilitate peer review, and provide educational value for new researchers in the field.

\section{Summary and Outlook}

In this work, we have released \textsc{QuRE}, the largest requirements quality dataset to date, with over 2,000 requirements from a single domain. The dataset has been labeled by up to three industry experts and has been used extensively in practice, including five years as part of a contractual setting, testifying to an industry-grade quality.
We provide a detailed description of the context and history of the dataset to enable future researchers to use it in a methodologically sound way.
Lastly, we compare the dataset with existing datasets, indicating that the dataset is linguistically consistent with other datasets. As a side effect of our analysis, we stumbled upon interesting discrepancies between synthetically generated and real-world requirements. This raises the question of whether analyzing synthetic data is a methodologically adequate use and how to prompt LLMs to generate requirements more similar to the existing dataset from industry. 
Finally, we discuss future research opportunities, but equally important, limitations for using the released data. We end with an encouragement to the community to finally assemble and define strategies and develop platforms for incrementally collecting and extending datasets in requirements research.

\bibliographystyle{IEEEtran}
\bibliography{sample}

\end{document}